\documentclass{aa}  

\usepackage{graphicx}
\usepackage{txfonts}
\usepackage[breaklinks=true]{hyperref} 
\hypersetup{final,colorlinks=true,linkcolor=blue,citecolor=blue,urlcolor=blue} 

\makeatletter
\renewcommand*\aa@pageof{, page \thepage{} of \pageref*{LastPage}}
\makeatother

\begin{document} 

\title{Simulations of radio-wave anisotropic scattering to interpret type III radio bursts measurements by Solar Orbiter, Parker Solar Probe, STEREO and Wind}

\titlerunning{Simulation of type III radio scattering and application to multi-spacecraft measurements}

\author{S. Musset \inst{1,} \inst{2}
  \and M. Maksimovic \inst{3} 
  \and E. Kontar \inst{2}
  \and V. Krupar \inst{4,5}
  \and N. Chrysaphi \inst{3,2}
  \and X. Bonnin \inst{3}
  \and A. Vecchio \inst{3,6}
  \and B. Cecconi \inst{3}
  \and A. Zaslavsky \inst{3}
  \and K. Issautier \inst{3}
  \and S. D. Bale \inst{7,8,9} 
  \and M. Pulupa \inst{7}}

\institute{European Space Agency (ESA), European Space Research and Technology Centre (ESTEC), Keplerlaan 1, 2201 AZ, Noordwijk, The Netherlands
    \and School of Physics \& Astronomy, University of Glasgow, Glasgow G12 8QQ, UK
    \and LESIA, Observatoire de Paris, Université PSL, CNRS, Sorbonne Université, Université de Paris, 5 place Jules Janssen, 92195 Meudon, France
    \and Goddard Planetary Heliophysics Institute, University of Maryland,Baltimore County, Baltimore, MD, USA
    \and Heliophysics Science Division, NASA Goddard Space Flight Center, Greenbelt, MD, USA
    \and Radboud Radio Lab, Department of Astrophysics/IMAPP - Radboud University, PO Box 9010, 6500 GL, Nijmegen, The Netherlands
    \and Space Sciences Laboratory, University of California, Berkeley, CA, USA
    \and Physics Department, University of California, Berkeley, CA, USA
    \and Stellar Scientific (now HELIOSPACE), 932 Parker St suite 2, Berkeley, CA, 94710, USA}

   \date{Received , 2021; accepted , 2021}


\abstract
{}
{We use multi-spacecraft observations of invididual type III radio bursts in order to calculate the directivity of the radio emission, to be compared to the results of ray-tracing simulations of the radio-wave propagation and probe the plasma properties of the inner heliosphere.}
{Ray-tracing simulations of radio-wave propagation with anisotropic scattering on density inhomogeneities are used to study the directivity of radio emissions. Simultaneous observations of type III radio bursts by four widely-separated spacecraft are used to calculate the directivity and position of the radio sources. The shape of the directivity pattern deduced for individual events is compared to the directivity pattern resulting from the ray-tracing simulations. }
{We show that simultaneous observations of type radio III bursts by 4 different probes provide the opportunity to estimate the radio source positions and the directivity of the radio emission. The shape of the directivity varies from one event to another, and is consistent with anisotropic scattering of the radio-waves.}
{}


\maketitle
\section{Introduction}

Solar type III radio bursts, which are produced by electron beams that are propagating along open magnetic field lines in the corona and interplanetary medium, are among the most intense radio sources in the kilometric range \citep{Wild1950, Wild1967}. Although, the precise mechanism of their production is not yet established, it is widely accepted that type III bursts originate via plasma emission mechanism in which the fast electrons form a bump-on-tail instability which then excites Langmuir waves at the local plasma frequency \citep{Ginzburg1958, Mel1980}. These Langmuir waves are then transformed into electromagnetic radiations by mechanisms which are still debated \citep{Stu1964,Zhe1970,Smi1976,Mel1987,Dul1985,Tkachenko_etal_2021}.


Radio-waves propagate through the solar corona and interplanetary medium where they are both refracted and scattered by turbulent plasma processes, with radio-wave scattering on density fluctuations being a dominant part of the propagation effects \citep{kontar_etal_2017, kuznetsov_etal_2020}. The observed properties of the radio sources are therefore a combination of their intrinsic properties at their emission site and these propagation effects. Radio-wave scattering causes both an increase of the radio source sizes and a very wide emission diagram responsible for detection of type III radio bursts at all angles in the heliosphere \citep{Steinberg1984,Steinberg1985,bonnin_etal_2008}, a shift of the radio source positions \citep{Fokker1965, chrysaphi_etal_2018}, and a widening of the intensity time profiles of the radio bursts \citep{Krupar2018, kontar_etal_2019}. 


The effect of radio-wave scattering on solar radio bursts has been studied using ray-tracing simulations \citep{steinberg_etal_1971,Thejappa2008,Krupar2018,krupar_etal_2020} where isotropic scattering is assumed. 
However, recent observational results strongly suggest that the properties of solar radio emissions cannot be successfully explained using isotropic scattering \citep{kontar_etal_2019,chen_etal_2020}. 
\cite{kontar_etal_2019} presented the results of ray-tracing simulations of radio-wave propagation in the heliosphere, assuming anisotropic scattering of the radio-waves. They showed that scattering was at the origin of the broadening of radio source sizes and radio burst time profiles, but that in average both properties could only be simultaneously described when anisotropic radio-wave scattering was considered.
   

The first measurements of the directivity of type III radio bursts were performed by \cite{caroubalos_steinberg_1974, caroubalos_etal_1974} using simultaneous observations from the Earth and from the probe Mars-3, at 169 MHz. At lower frequencies (100-500 kHz), \cite{Dul1985} and \cite{lecacheux_etal_1989} reported that type III radio bursts were detected by instruments irrespective to the position of the radio source. The first stereoscopic directivity measurements were reported by \cite{hoang_etal_1997} who used a combination of ground-based observations from the ARTEMIS spectrograph around 150 MHz and space observations from the radio receiver on board Ulysses (up to 1 MHz). The first stereoscopic measurements in the same frequency range to study the radio burst directivity were performed by \cite{bonnin_etal_2008} using observations from the Ulysses and Wind spacecraft. \cite{bonnin_etal_2008} used the observations of more than 2000 radio bursts observed simultaneously by both probes, to statistically derive the directivity of the type III radio bursts between 80 and 1000 kHz.  They confirmed the frequency dependence of the averaged radio burst directivity that was observed by \cite{hoang_etal_1997}. These studies had to rely on a statistical analysis of a large sample of bursts as they had access to only two simultaneous observations of a single event: the directivity profile is therefore calculated using hundreds to thousands of radio flux ratios. With the launch of Solar Orbiter \citep{muller_etal_2020} and Parker Solar Probe \citep[PSP, ][]{fox_etal_2016}, radio measurements at two different points in the heliosphere can be added to the already existing measurements of STEREO-A \citep[Solar Terrestrial Relations Observatory,][]{kaiser_Stereo} and Wind \citep{ogilvie_wind}, providing for the first time the opportunity to study radio emission directivity for single type III radio bursts. In this paper, we present the analysis of 5 events observed in the early phase of the Solar Orbiter mission (July and November 2020). These observational measurements are compared to the predictions of ray-tracing simulations of the radio-wave propagation with anisotropic radio-wave scattering on turbulent fluctuations of the ambient plasma density. The ray-tracing simulation results are presented in section \ref{sec:simu}, and the observations are presented in section \ref{sec:obs}. The comparison between the simulation results and the observed properties of the type III bursts, and the implication regarding the radio emission directivity and the properties of the ambient plasma, is presented in section \ref{sec:discussion}.


\section{Directivity of radio emission from radio-wave propagation simulations}
\label{sec:simu}

\subsection{Simulation setup}

We perform simulations of radio-wave propagation in the heliosphere using the ray-tracing simulations described in \cite{kontar_etal_2019}. 
The ray-tracing simulations describe the propagation of photons through the interplanetary medium, where anisotropic density fluctuations (and thus anisotropic scattering) can be assumed. 
A radio source is placed at a given distance from the solar surface, and the corresponding plasma frequency is deduced from the position of the source, assuming a density model.
We used the density model described in \cite{kontar_etal_2019}:
\begin{equation}
n(r) = 4.8 \times 10^9 \left(\frac{R_s}{r}\right)^{14} + 3 \times 10^8 \left(\frac{R_s}{r}\right)^{6} + 1.4 \times 10^6 \left(\frac{R_s}{r}\right)^{2.3}
\label{eqn:densitymodel}
\end{equation}
were $n(r)$ is the plasma density in cm$^{-3}$, $r$ the position of the source and $R_s$ is the solar radius.

The radio frequency is usually set to a factor of 1.1 or 1.2 (for fundamental emission) or 2 (harmonic emission) of the plasma density, and stays constant during the radio-wave propagation. 

During their propagation, radio-waves are subject to scattering due to turbulent fluctuations of the ambient plasma density. An anisotropic spectrum of density fluctuation $S(\mathbf{q})$ ($\mathbf{q}$ being the wave-vector of electron density fluctuations) is assumed, with an axial symmetry. The spectrum is thus parameterised as:
\begin{equation}
S(\mathbf{q}) = S \left( \sqrt{q_{\bot}^2 + \alpha^{-2} q_{\parallel}^2} \right)
\end{equation}
where $\alpha$ is the anisotropy factor defined as the ratio of perpendicular to parallel correlation lengths: $\alpha = h_{\bot}/h_{\parallel}$. When $\alpha << 1$, the spectrum of density fluctuations is dominated by the fluctuations in the perpendicular direction. In the case of isotropic scattering, $\alpha = 1$.


The photon propagation is simulated until photons reach a distance from the Sun where both refraction and scattering become negligible(i.e. when the photon frequency, which is kept constant in the simulations, becomes much larger than the local plasma frequency), or until they reach a distance of 1 au. For each photon, at the end of the simulation, the time $t$ of arrival and its position $\mathbf{r}$ and wave-vector $\textbf{k}$ are recorded, in a Sun-centered coordinate system.


\subsection{Anisotropic scattering and directivity}
\label{sec:directivity-anisotropy}

\begin{figure}
\includegraphics[width=0.8\linewidth]{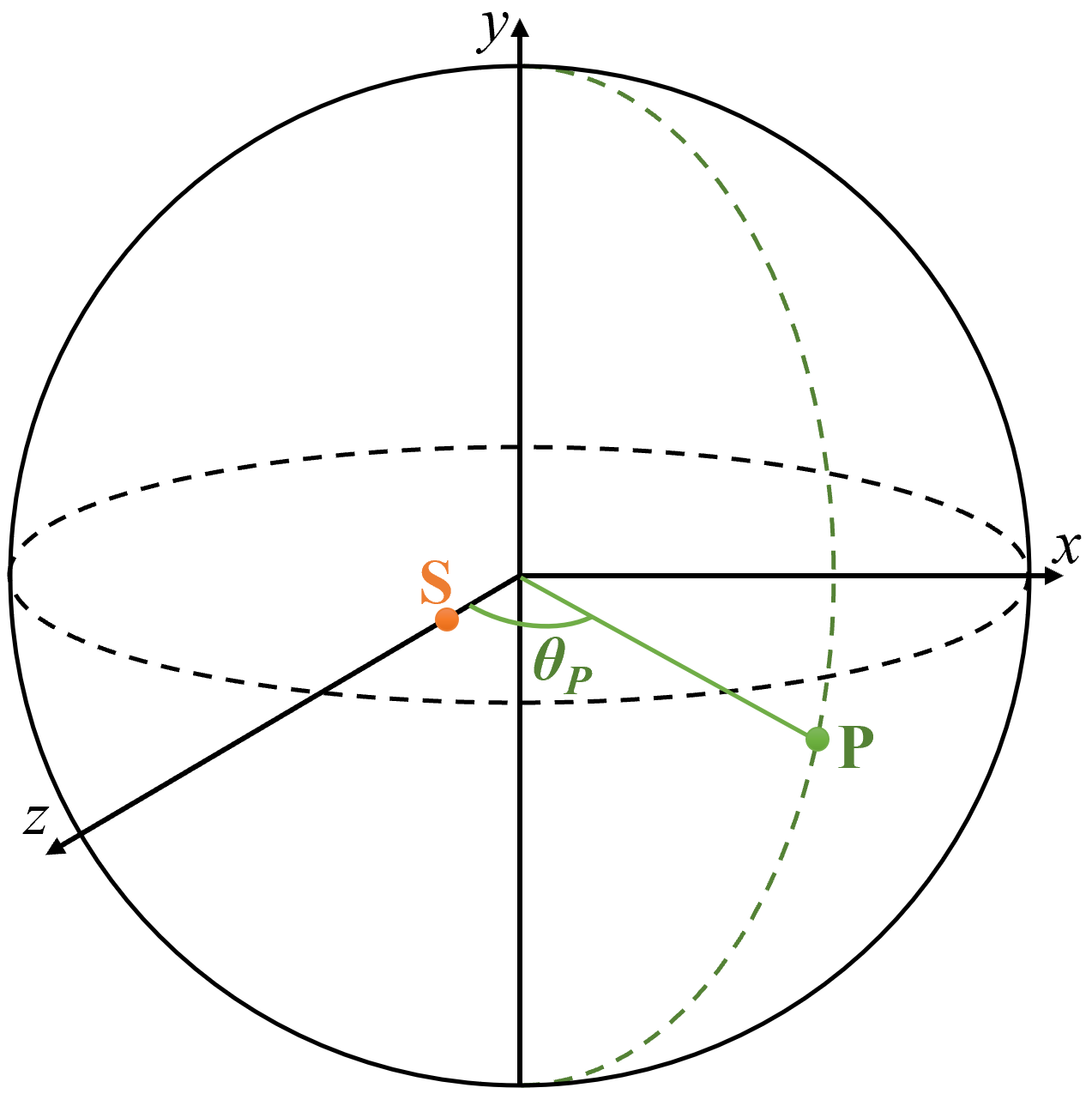}
\caption{Geometry of the simulation: the radio source $S$ is emitted on the $z$-axis. The angle $\theta_P$ is the angle between the direction of the radio source $S$ (the $z$-axis) and the direction of an observer at position $P$. In this paper, the radio source is always situated on the $z$-axis while the probe $P$ have different positions in the 3-D space.}
\label{geometry}
\end{figure}

The geometry in which the results of the ray-tracing simulation are analysed is summarized in figure \ref{geometry}. The direction of propagation of the radio source, the $z$-axis, is a symmetry axis for the directivity of the radio emission. Therefore, the directivity can be described as a function of the angle $\theta_P$ defined in the figure, or of its cosine, $\mu = \cos (\theta_P)$.
In our study, as in past studies \citep[e.g.][]{hoang_etal_1997,bonnin_etal_2008}, we assume that the directivity follows an exponential shape as a function of $\mu$:
\begin{equation}
F(\mu) = C_0 \exp \left( - \frac{\left( 1-\mu \right)}{\Delta \mu} \right)
\label{directivity_model}
\end{equation}
where $C_0$ is a normalization constant and $\Delta \mu$ the key parameter controlling the shape of the radio emission directivity pattern.
In the simulations, the parameter $\mu$ is calculated as $\mu = k_z/\left|\textbf{k}\right|$.
 
The radio emission directivity pattern depends on the anisotropy factor $\alpha$. We remind here that $\alpha = 1$ for isotropic scattering, and that $\alpha < 1$ results in a stronger scattering in the perpendicular direction. Smaller values of $\alpha$ should therefore lead to a more peaked directivity \citep{kontar_etal_2019}, which translates to smaller values of the parameter $\Delta \mu$ in equation \ref{directivity_model}.
We performed ray-tracing simulations of the radio-wave propagation with the following settings:
\begin{itemize}
\item A radio source located at 11 R$_S$, leading to a plasma density $f_{pe} = 681 $ kHz, and fundamental radio emission emitted at $1.1 \times f_{pe}$ \footnote{We also simulated a radio sources at a frequency of $1.2 \times f_{pe}$ and obtained the same directivity profiles. }
\item Anisotropy factor $\alpha = 0.25, 0.30, 0.40, 0.50, 0.60, 0.75, 1.00$
\end{itemize}

\begin{figure}
\includegraphics[width=\linewidth]{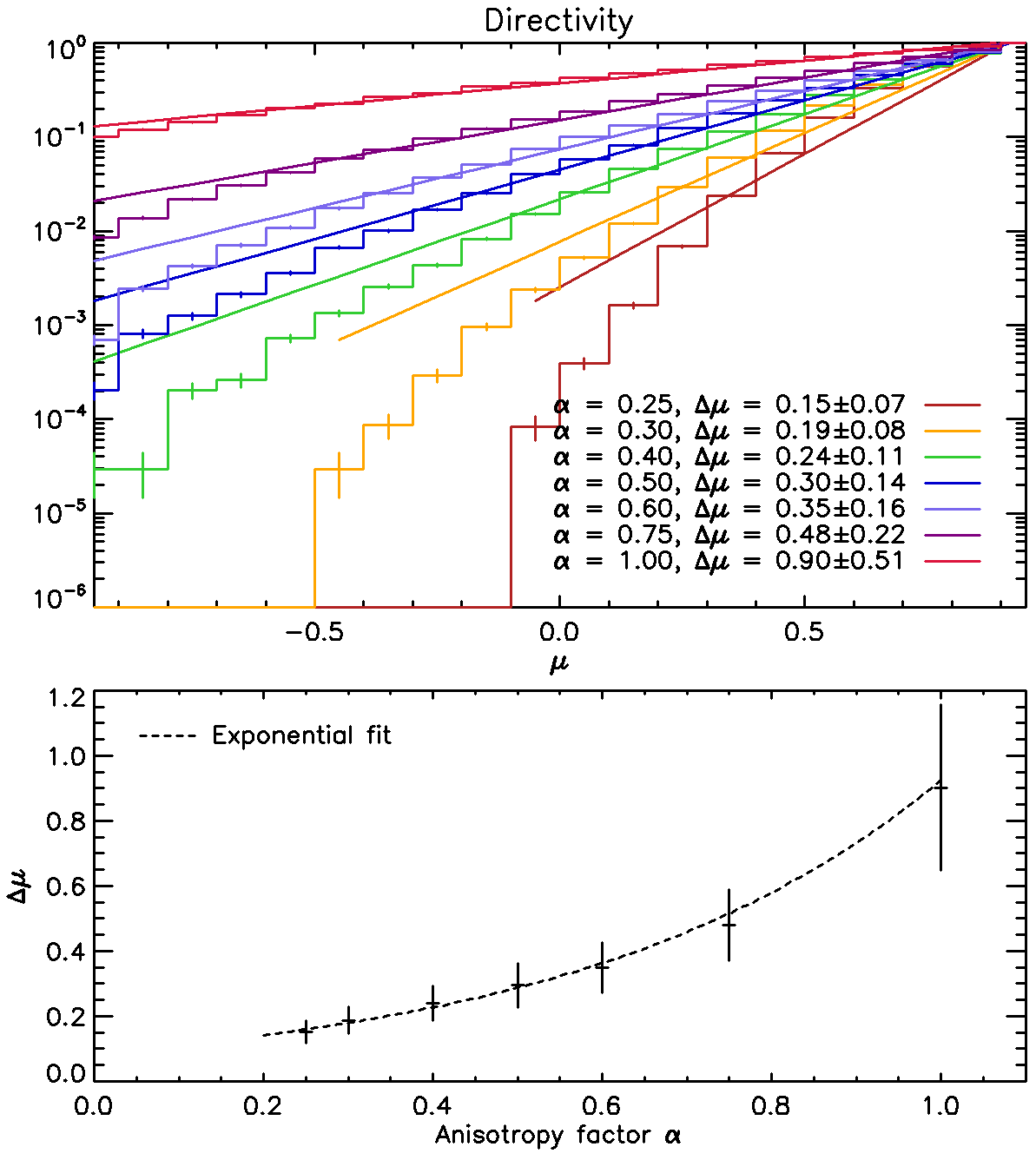}
\caption{Top: normalized directivity as histograms calculated from the results of the simulations described in section \ref{sec:directivity-anisotropy}, with anisotropy factors $\alpha =$ 0.25, 0.30, 0.40, 0.50, 0.60, 0.75 and 1.0. The result of the exponential fit is shown as a line for each curve, and the values of the parameter $\Delta \mu$ resulting from this fit are displayed in the low right corner. Bottom: Evolution of the parameter $\Delta \mu$ with the anisotropy factor $\alpha$. The distribution is fitted with an exponential model: $\Delta \mu \propto e^{(a_0 \alpha)}$, which gives $a_0 = 2.2 \pm 0.3$. The result of this fit is shown as a dashed line.}
\label{directivity_anisotropy}
\end{figure}

The directivity obtained at these frequencies for photons arriving around the peak time of the corresponding lightcurve is shown in figure \ref{directivity-freq}. To select the photons around the peak time, we generated the lightcurves for the radio emission using the propagation time of each photon resulting from the ray-tracing simulations, and selected photons arriving in the time interval where the lightcurve is above 80 \% of the peak time intensity.
The directivity obtained with these different simulations, and normalized to the maximum intensity, is displayed in figure \ref{directivity_anisotropy}. Each directivity curve was fitted with the model described by equation \ref{directivity_model}, using the \texttt{mpfitfun} procedure which performs Levenberg-Marquardt least-squares minimization \citep{mpfit_idl}. The result of this fit is also shown in figure \ref{directivity_anisotropy}. As seen in the figure, 
in the case of isotropic scattering ($\alpha=1$), the exponential model is a good approximation of the directivity. When we increase the anisotropy of the scattering (by decreasing the ratio $\alpha$), the directivity starts to deviate from an exponential shape. However, we keep this model in order to quantify the change of shape of the directivity with the anisotropy factor. As we increase the anisotropy, the directivity distribution gets thinner, as expected, leading to smaller values of the parameter $\Delta \mu$ of the exponential model. In the bottom panel of figure \ref{directivity_anisotropy}, the evolution of $\Delta \mu$ resulting from the fits, as a function of the anisotropy factor $\alpha$, is displayed. This evolution is fitted by a model of the form $\Delta \mu \propto e^{(a_0 \alpha)}$, which gives $a_0 = 2.2 \pm 0.3$. 

\subsection{Evolution of directivity with frequency}
\label{sec:directivity-freq}

\begin{figure}
\includegraphics[width=\linewidth]{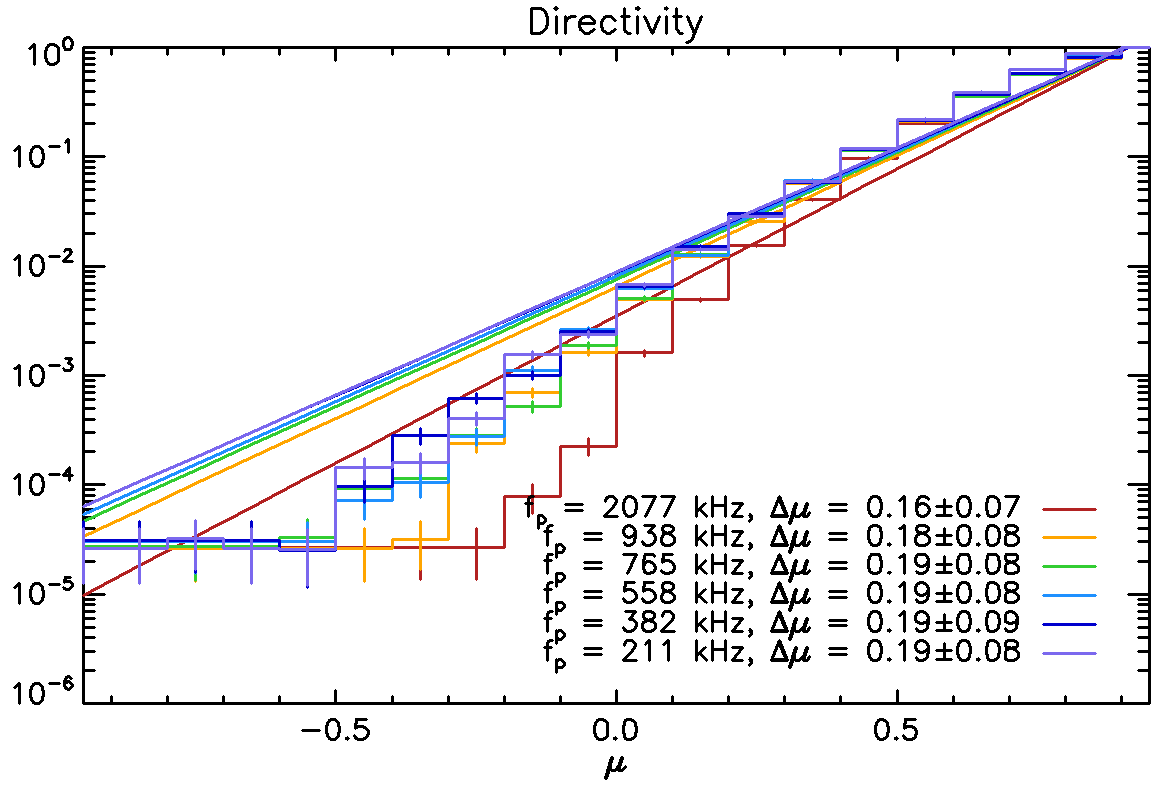}
\caption{Normalized directivity as histograms calculated from the results of the simulations described in section \ref{sec:directivity-freq}, with anisotropy factor $\alpha =$0.30 and plasma frequency of 2077, 938, 765, 558, 382 and 211 kHz in red, orange, green, light blue, dark blue and purple, respectively. The result of the exponential fit is shown as a line for each curve, and the values of the parameter $\Delta \mu$ resulting from this fit are displayed in the low right corner.}
\label{directivity-freq}
\end{figure}

The influence of frequency on the directivity of radio bursts in the radio-wave propagation simulations is examined with the following simulation settings:
\begin{itemize}
\item A radio source located at 5, 8.5, 10, 13, 18 or 30 R$_S$, leading to a plasma density $f_{pe} = 2077, 938, 765, 558, 382 $ or 211 kHz respectively, and fundamental radio emission emitted at $1.2 \times f_{pe}$
\item Anisotropy factor $\alpha =  0.30$
\end{itemize}

As described in section \ref{sec:directivity-anisotropy}, the directivity shown in  figure \ref{directivity-freq} is obtained here for photons arriving around the peak time. As can be seen in the figure, there is small dependence of the shape of directivity on frequency, with the parameter $\Delta \mu$ decreasing with increasing frequency. We note that this evolution is a small and remains within the error bars for $\Delta \mu$. We also note that when we select all photons (instead of the photons arriving near the peak time of the lightcurve), the directivity pattern does not depend on the frequency anymore.

\subsection{Time profiles at different angles}

Radio-wave scattering is responsible for delays in the propagation time and results in a broadening of the time profiles of type III radio bursts. We used the simulations presented in section \ref{sec:directivity-anisotropy} to quantify the time profile broadening introduced by scattering, and the effect of both the radio emission frequency and the anisotropy of the scattering process on the time profile widths (or decay times). 

\begin{figure}
\includegraphics[width=0.95\linewidth]{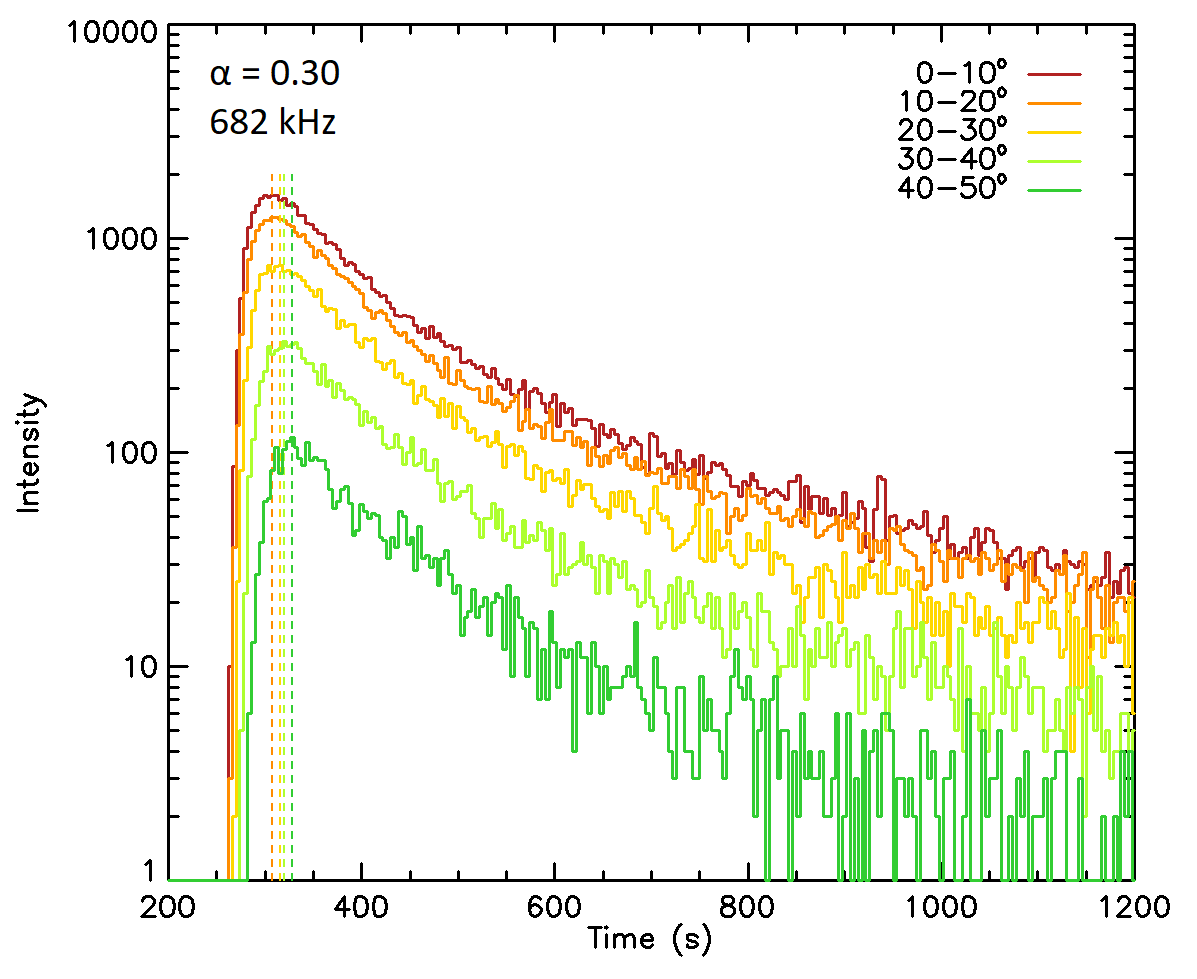}
\caption{Example of time profiles obtained for photons collected on a sphere of radius 111 $R_S$ (where propagation effects become negligible), at different angles $\theta_P$, from the simulations of radio-wave propagation from an emission source at 11 $R_S$ and an anisotropy factor $\alpha = 0.30$. Dashed vertical lines indicate the peak time for each time profile.}
\label{timeprofiles}
\end{figure}

\begin{figure}
\includegraphics[width=0.9\linewidth]{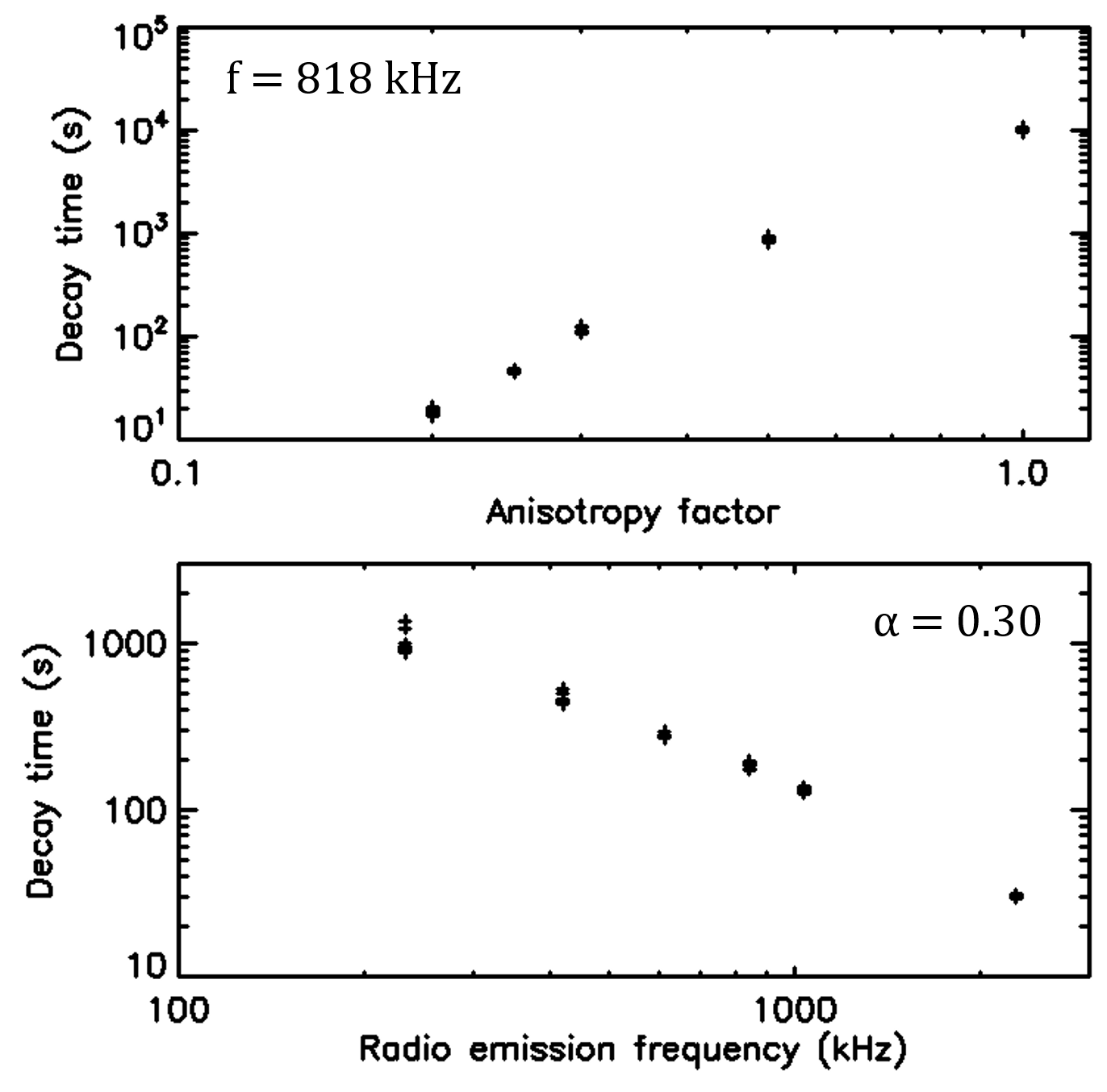}
\caption{Evolution of the lightcurve decay times in the simulations. Top: evolution of the decay time as a function of the anisotropy factor $\alpha$, for emissions at 818 kHz. Bottom: evolution of the decay time as a function of the radio emission frequency (assuming fundamental emission), for an anisotropy factor of 0.3. On each plot, the decay time is determined at different angles from the radio source.}
\label{decaytime}
\end{figure}

An example of time profiles collected in different directions in space (i.e., at different angles from the radio source direction, the $z$-axis) is shown in figure \ref{timeprofiles}. 
As it can be seen in the figure, the time profiles cannot be fully described by an exponential decay: there is a break in the slope happening roughly when the intensity falls below 10\% of the maximum intensity. This behaviour is introduced by the anisotropy in the scattering process, which is responsible for an "echo" of the lightcurve, as described using the same ray-tracing simulations, but at higher frequencies, by \cite{kuznetsov_etal_2020}.

For each time profile, the peak time and the full width at half maximum (FWHM) is determined. 
Then the decay part of the time profiles is fitted with an exponential model of the form:
\begin{equation}
    P = P_0 \exp{ \left( - \frac{t-t_0}{\tau} \right)}
\label{eqn:timeprofile}
\end{equation}
where $\tau$ is the decay time. This fit is performed using the \texttt{mpfitfun} procedure which performs Levenberg-Marquardt least-squares minimization \citep{mpfit_idl}. For this fit, we used only the part of the profile after the peak time, for which the flux is above 10\% of the peak flux. 

The evolution of the decay time with the anisotropy factor and the radio frequency is shown in figure \ref{decaytime}. This analysis demonstrates that:
\begin{itemize}
    \item the decay time does not vary significantly from one point of view in space to another (from one angle to another)
    \item the decay time varies significantly with the anisotropy factor $\alpha$: as shown in the top panel of figure \ref{decaytime}, the decay time increases significantly as the anistropy factor increases towards $\alpha = 1$, as reported in \cite{kontar_etal_2019}
    \item  the decay times varies significantly with the radio emission frequency, as shown in the bottom panel of figure \ref{decaytime}: the decay time decreases as the frequency increases, also as reported in \cite{kontar_etal_2019}.
\end{itemize}

\begin{figure*}
\includegraphics[width=0.95\linewidth]{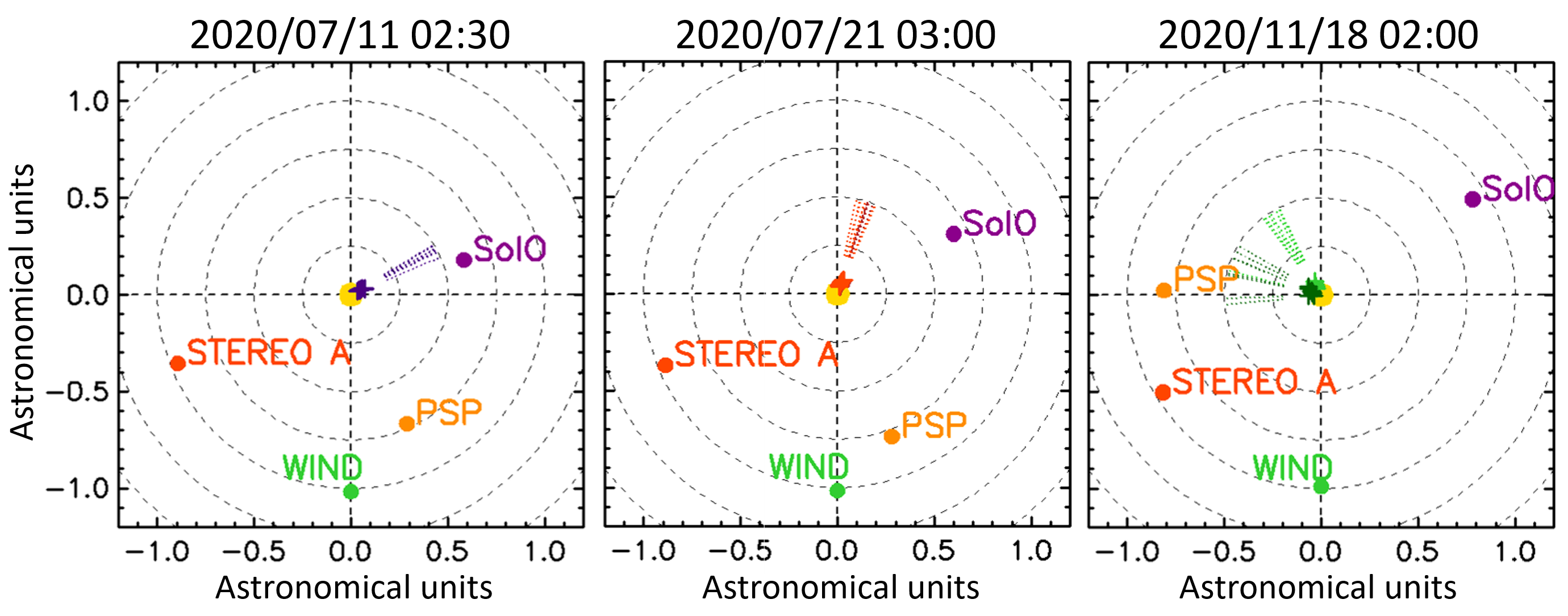}
\caption{Position of Solar Orbiter, Parker Solar Probe, STEREO-A and Wind projected in the plane of the Heliocentric Earth Ecliptic (HEE) coordinate system. Left: position on July 11 2020 at 02:30 UT. The position of the radio source is shown as crosses near the Sun. Middle: position on July 21 2020 at 03:00 UT. The positions of the radio sources from the two events of that day are indicated by coloured crosses near the Sun. Since both events are from the same region, they are superimposed. Right: position on November 18 2020 at 02:00 UT. The position of the radio sources for the two events of that day are shown as coloured crosses near the Sun. On this day there is a difference between the location of the radio sources from the first event and from the second event. The determination of the radio source location is described in the text. On each plot, dotted lines are used to point the direction of the radio sources}
\label{figure_position}
\end{figure*}


\section{Multi-spacecraft observations of type III radio bursts}
\label{sec:obs}


In this section we describe briefly the space-born radio instruments which we have used in this study and the corresponding data which have all been calibrated to radio Solar Flux Units (SFU, 1 SFU=$10^{-22}W/m^2/Hz$). All these instruments operate in the radio kilometric range, with corresponding frequencies spanning between a few kHz to 10-20 MHz, just below the ionospheric radio cutoff. Then we describe our observations and analyses results.

\subsection{Descriptions of the instruments}

We have used radio observations embarked on board the Wind \citep{ogilvie_wind}, the STEREO-A \citep{kaiser_Stereo}, the Parker Solar Probe \citep{fox_etal_2016}, and the Solar Orbiter \citep{muller_etal_2020,zouganelis_etal_2020} spacecraft.

On Wind, the Waves experiment \citep{bougeret_etal_1995} has three electric dipole antennas; two of them are co-planar and orthogonal wire dipole antennas in the spin-plane, whereas, the other is a rigid spin-axis dipole. 
We use Wind/Waves demodulated data, derived from a so-called direction-finding method by \cite{manning_fainberg_1980} and calibrated using the radio galaxy background \citep{Zaslavsky_etal_2011}.
This technique provides the absolute radio flux of the observed sources, their polarization properties and their directions and sizes.

On STEREO-A, the S/Waves instrument \citep{bougeret_etal_2008} performs radio measurements using three orthogonal antenna monopoles. This time the calibration is based on direction finding techniques specific for three-axis stabilized spacecraft \citep{cecconi_etal_2008,krupar_etal_2012,krupar_etal_2016}). 

On Parker Solar Probe, we use data recorded by the Radio Frequency Spectrometer \citep[RFS, ][]{Pulupa,psp_fields_rfs_data} connected to the FIELDS electric antennas \citep{Bale_Fields}. On PSP, the full direction finding radio data pipeline is still in development. However for sources observed in the frequency range we use in this study, which is from 400 to 1000 kHz, the heliospheric locations (12 to 6 $R_s$ using the \cite{Leblanc} density model) are such that the angles between their k-vectors and the radial direction is negligible. Under these conditions the radio flux can be defined as the sum of the power spectral densities measured by each of the two crossed dipoles, multiplied by their respective $sin(\theta_{Di})^2$, $\theta_{Di}$ being the angle between the radial and the given FIELDS dipole. For the events studied here, $\theta_{Di}$ range between 66 and 81$^{\circ}$ The conversion from power spectral densities to radio fluxes in SFU is obtained using the gains as defined by \cite{Maksimovic_psp}.

Finally on Solar Orbiter, we use radio data recorder by the radio receiver part of the RPW (Radio and Plasma Waves) instrument \citep{Maksimovic_RPW}. As for PSP, the full direction finding radio data pipeline is also in development for RPW. Therefore we make the same assumptions for the directions of the radio sources k-vectors. The only difference is that, contrary to PSP, the RPW antenna dipoles are always perpendicular to the radial direction. We therefore define the radio flux for SO as being two times the power spectral density as measured by one dipole (the factor 2 comes from the fact that we assume the source to be non-polarized), multiplied by the calibration gains extensively described by \cite{Vecchio}.

\subsection{Observations}

\begin{figure}
\includegraphics[width=0.95\linewidth]{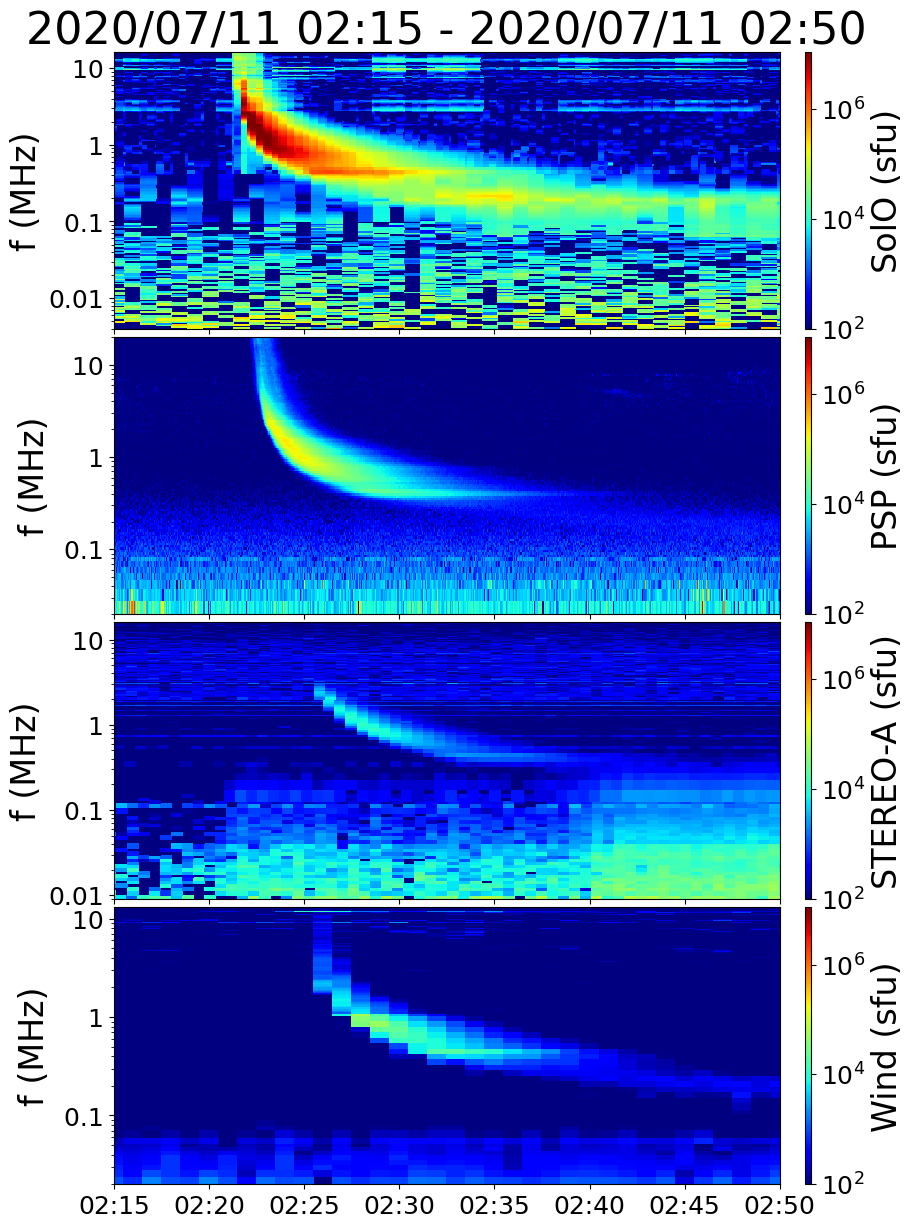}
\caption{Dynamic spectrum of the radio fluxes in SFU for the four spacecraft for the event on July 11 2020 around 02:30 UT; from top to bottom: Solar Orbiter, Parker Solar Probe, STEREO-A, Wind.}
\label{dyna_spectrum}
\end{figure}

For this study we selected five type III radio bursts that were observed by four widely-separated spacecraft: the first event was observed on July 11 2020 around 02:30 UT, the second and third events were observed on July 21 2020 around 03:00 and 07:00 UT respectively, and the fourth and fifth events were observed on November 18 2020 around 02:00 and 22:30 UT respectively. The positions of the probes on each day \textcolor{blue}{are} displayed in figure \ref{figure_position}. Figure \ref{dyna_spectrum} displays dynamic spectra of the radio fluxes in SFU for the four spacecraft for the event on July 11 2020 around 02:30 UT. As for the other events we have chosen to analyse well defined and isolated type III bursts, clearly visible on the four probes.

We selected 8 frequency channels in which to analyse the radio emission: 411, 430, 511, 635, 663, 755, 788 and 979 kHz. These particular frequencies have been chosen because they correspond to clean RPW frequencies which are not polluted by the strong electromagnetic emission from the Solar orbiter platform \citep{Maksimovic1st}

\begin{figure}
\includegraphics[width=0.98\linewidth]{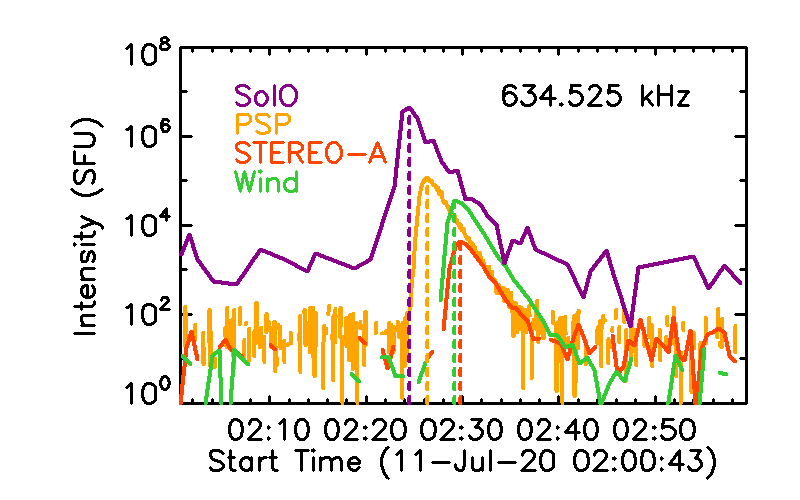}
\caption{Light curves of the radio fluxes at 634.5 kHz for the event on July 11 2020, observed at Solar Orbiter, Parker Solar Probe, STEREO-A and Wind. For each of these light curves, the median of the radio signal on an interval of 30 minutes before 2:20 UT has been removed. For each of the probes, the time of the peak flux is indicated by the vertical dashed lines. }
\label{light_curves}
\end{figure}

Figure \ref{light_curves} displays temporal variations  of the radio fluxes at 634.5 kHz, for the four spacecraft and for the event on July 11 2020 around 02:30 UT. For each of these light curves, the median of the radio signal on an interval of 30 minutes before 2:20 UT has been removed. All remaining light curves have a classical shape, rapidly reaching a maximum and decaying exponentially. For each of the probes, the time of the peak flux is indicated by the vertical dashed lines.

\subsubsection{Decay times of the Type III radio fluxes}

\begin{figure}
\includegraphics[width=0.98\linewidth]{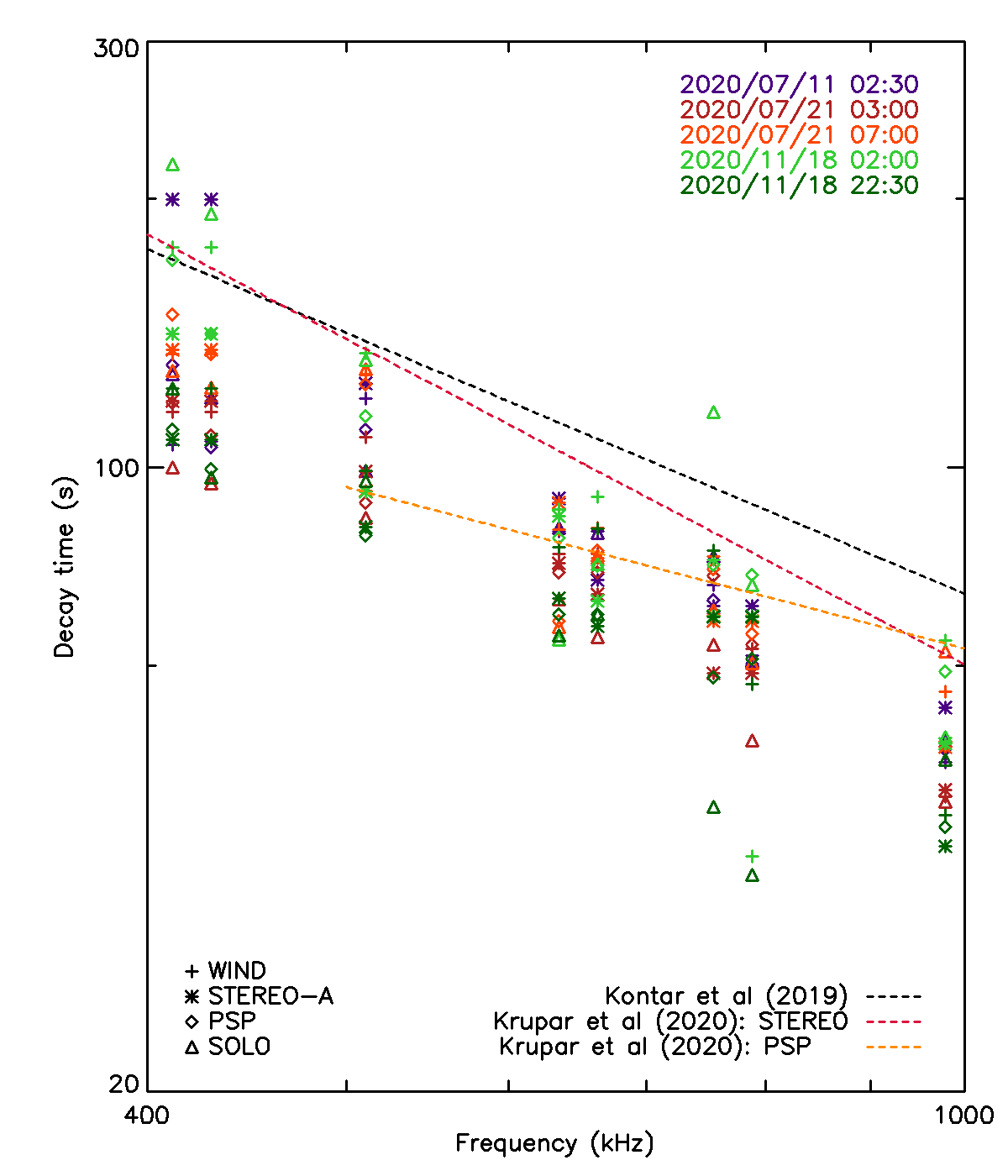}
\caption{Decay times of the radio bursts presented as a function of the frequency of the radio emission. Different colours show different type III bursts, and different symbols show the different probes. The dashed black line show the result of the fit to the data performed by \cite{kontar_etal_2019} on several data sets. The dashed red and orange lines show the result of the fit on the distribution of decay times from 30 radio bursts observed by STEREO and Parker Solar Probe respectively \citep{krupar_etal_2020}}
\label{observation_decaytime}
\end{figure}

The decay time of the radio bursts was determined at each of the 8 selected frequencies by fitting an exponential curve to the decaying part of the burst, using the same model as the one used to analyse the lightcurves from the simulations (equation \ref{eqn:timeprofile}). The interval chosen to perform the fit is from the peak time to the time when the flux falls below 10 \% of the peak intensity.
 The result of these fits is shown in figure \ref{observation_decaytime}. As expected, the decay time decreases with increasing frequency. However, there is no clear variation of the decay times between the different events. These observed decay times align with previous measurements of the evolution of the decay time with frequency, such as the compilation of measurement fitted in \cite{kontar_etal_2019}, or the recent observations of 30 type III bursts by STEREO and Parker Solar Probe presented in \cite{krupar_etal_2020}. 

\subsubsection{Directivity fit to observations}

\begin{figure}
\includegraphics[width=0.8\linewidth]{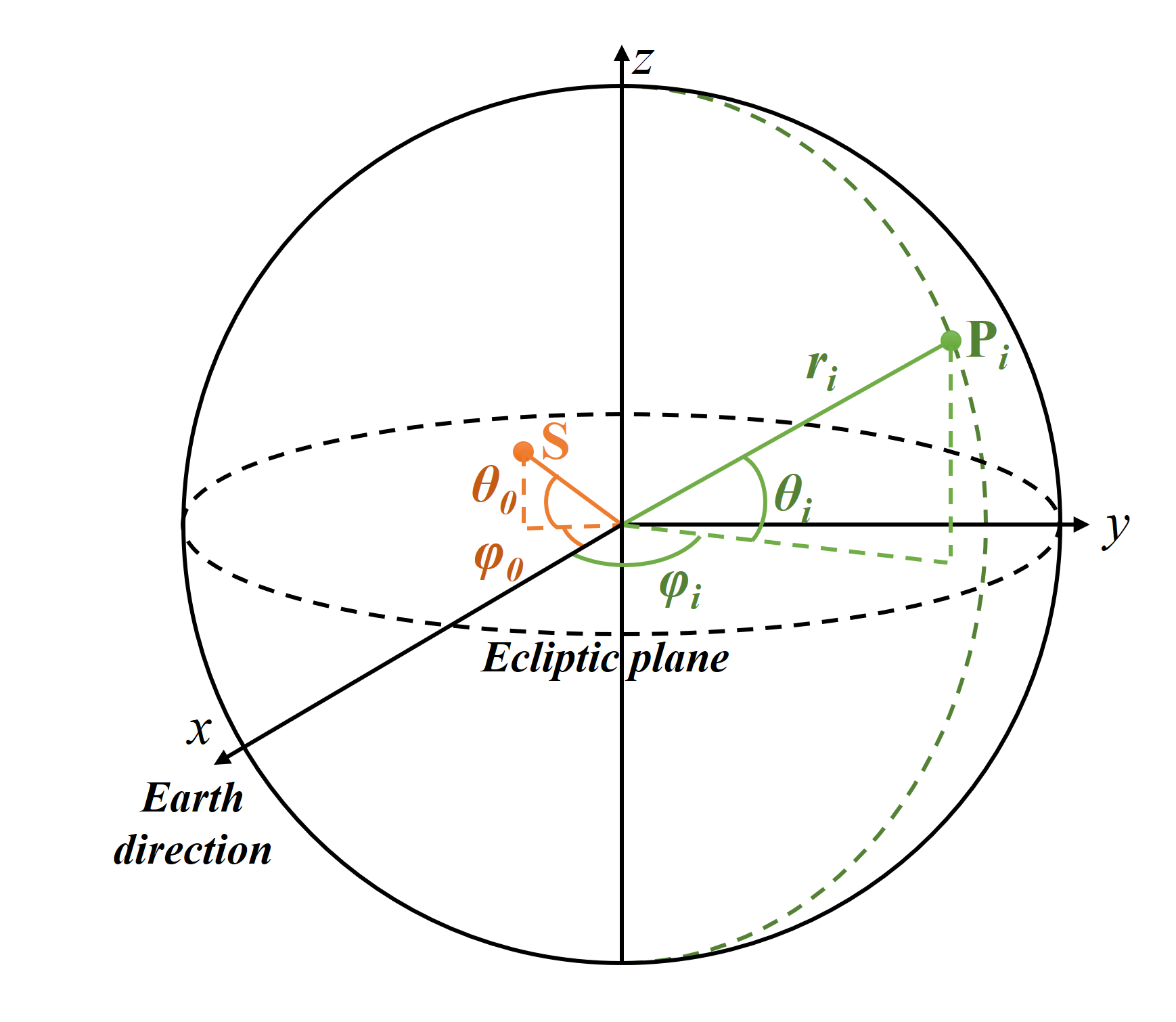}
\caption{Position of the radio source and probe $i$ in the HEE coordinate system: longitudes are noted with angle $\varphi$, latitudes with angle $\theta$.}
\label{geometry_obs}
\end{figure}

Simultaneous observations of type III radio bursts with 4 different observation angles allow a determination of the directivity of the radio emission. While the position of each instrument is well known, the position of the radio source is an unknown parameter. We therefore used the radio flux measurements at the four spacecraft to determine both the shape of the radio emission directivity pattern and the location of the radio source. The geometrical used is illustrated in figure \ref{geometry_obs}: the positions of the probes are given in the Heliocentric Earth Ecliptic (HEE) coordinate system, where the $z$-axis is the solar rotation axis and the $x$-axis is in the plane containing the $z$-axis and Earth: the coordinates $(r_i, \varphi_i, \theta_i)$ correspond to the distance to the Sun, the longitude and latitude of probe $i$ in this coordinate system.

During our observations, the different spacecraft are at different distances from the radio source. Since the radio source remains very close to the Sun, we assume that we can correct the flux intensity of each spacecraft for their relative distance to the radio source by using the heliocentric distance of the spacecraft. The fluxes corrected for the different distances are noted $I_{au}$ as they represent the radio flux that would have been measured if the spacecraft was positioned at 1 au from the Sun.

The radio flux at 1 au depends on the position of the observation point with respect to the radio source, the intensity of the radio emission, and the directivity of the emission. Therefore, it can be described as a function of the source position $(\varphi_0, \theta_0)$, the probe position $(\varphi, \theta)$, and the parameter $\Delta \mu$ for the directivity. The directivity is modelled using equation \ref{directivity_model} with $\mu = \cos (\varphi - \varphi_0) \cos (\theta - \theta_0)$ and $C_0$ the maximum intensity, when $\varphi = \varphi_0$ and $\theta = \theta_0$.

In the case of multi-spacecraft observations of type III radio bursts, the position of the probes $(\varphi_i, \theta_i)$ is known for each probe $i$, while the position of the source $(\varphi_0, \theta_0)$, and the parameters $\Delta \mu$ and $C_0$ are the four unknowns. Using the \texttt{MPFIT} procedure, we fit a directivity profile to best represent the radio fluxes measured at the different spacecraft. The result of the fit gives both a measurement of the directivity parameter $\Delta \mu$ and a location of the radio source $(\varphi_0, \theta_0)$, independent of other methods to locate the source (like triangulation).

\begin{figure}
\includegraphics[width=0.8\linewidth]{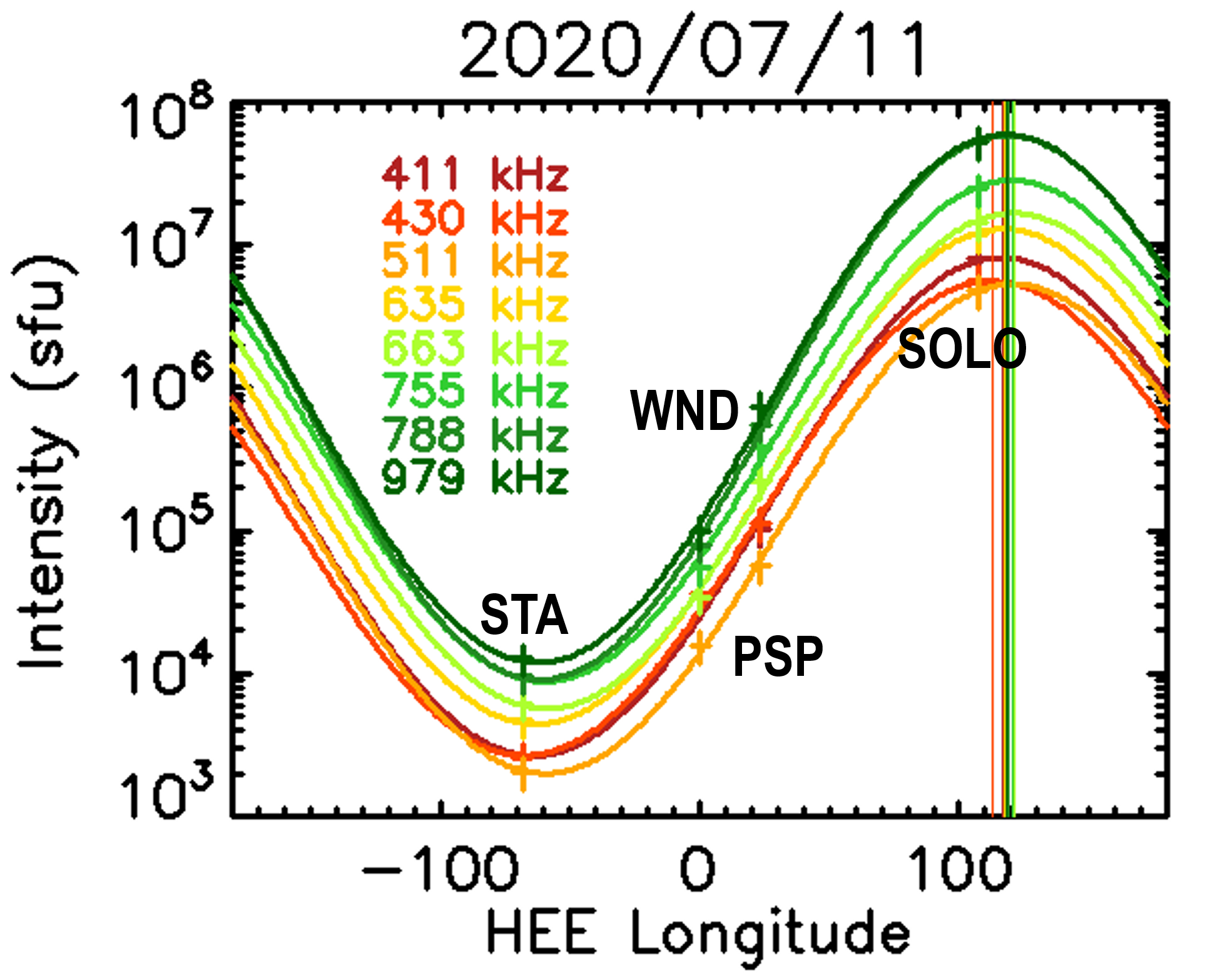}
\caption{Directivity fit on the peak fluxes measured on the 2020/07/11 event at different frequencies indicated with different colours. STEREO-A, WIND, Parker Solar Probe and Solar Orbiter measurements are indicated by crosses and labelled as STA, WND, PSP and SOLO respectively. The vertical lines show the position of the radio source in longitude as determined from the directivity fit.}
\label{fit_result}
\end{figure}

For the events presented in this paper, the probes are all located close to the ecliptic plane, at low latitudes: therefore, there are poor constraints on the latitude of the radio source $\theta_0$, and this parameter was kept fixed to 0 for the directivity fit.
The directivity model is fitted using the peak intensity at each of the 8 frequencies selected, resulting in 8 directivity profiles for each event. An example of the directivity fit shown in the ecliptic plane is displayed in figure \ref{fit_result}. The main results of these fits are the amplitude of the radio emission directivity pattern of the radio burst, characterized by the parameter $\Delta \mu$, and the position of the radio source (longitude, with the assumption of a source in the ecliptic plane).  

Changes in the level of uncertainties in the measured fluxes at the different probes do not significantly affect the results of the directivity fit: while the uncertainty on the parameters is linked to the uncertainty on the measured fluxes, the parameters' values themselves do not significantly change. Moreover, even if the fluxes from the RPW and FIELDS are not determined with the full direction finding radio data pipelines in this paper, no significant change to the results presented here should be expected once the final calibration to the data is applied.

In order to determine the radio source location, we use the results of our directivity fit, which provide the longitude and latitude of the source, and estimate the radial distance of the source to the Sun using a density model. In this paper, the latitude of the source is assumed to be 0, and we used the density model described by equation \ref{eqn:densitymodel}, assuming that the radio emission is emitted at the fundamental. The resulting source positions are displayed in each panel of figure \ref{figure_position} as coloured crosses for each event. On July 21 2020, the radio sources positions from the two type III radio bursts overlap. On November 18 2020, the radio sources positions from the two type III radio bursts are slightly shifted, it is likely that these sources are emitted by electron beams originating from the same active region. 
For all events, the deduced locations of the radio emission sources roughly correspond to a longitude at which active regions are present in the synoptic maps of the solar surface's magnetic field.

\begin{figure}
\includegraphics[width=0.98\linewidth]{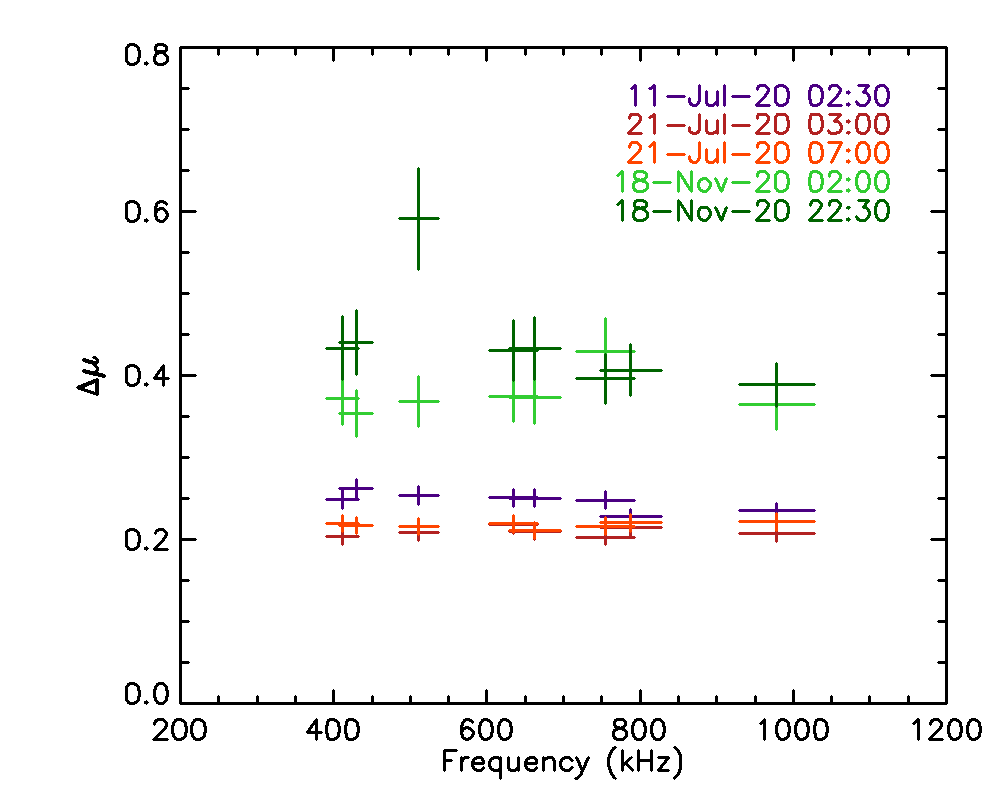}
\caption{Parameter $\Delta \mu$ determined during the directivity fitting of the peak fluxes for the 5 radio bursts selected.}
\label{observation_deltamu}
\end{figure}

\begin{table}[]
    \centering
    \begin{tabular}{|c|c|c|}
    \hline
         event            & Mean value of $\Delta \mu$ & Standard deviation \\
    \hline
         2020/07/11 02:30 &  0.25    &  0.01   \\
         2020/07/21 03:00 &  0.21    &  0.01    \\
         2020/07/21 07:00 &  0.22    &  0.01    \\
         2020/11/18 02:00 &  0.4     &  0.2    \\
         2020/11/18 22:30 &  0.4     &  0.4    \\
    \hline
    \end{tabular}
    \caption{$\Delta \mu$ averaged over the different frequencies for each event, and corresponding standard deviation.\protect\footnotemark}
    \label{tab:deltamu}
\end{table}
\footnotetext{The standard deviation here is calculated over the different frequencies for each individual event. }

The amplitude of the radio burst directivity, determined from the directivity fit, is displayed as a function of frequency in figure \ref{observation_deltamu}. As seen in the figure, the values of the parameter $\Delta \mu$ can vary significantly from one event to another. However, no clear variation with frequency is found for the parameter $\Delta mu$. The mean values of the parameters and the standard deviation, for each event, are displayed in \ref{tab:deltamu}. For the first three events, the value of $\Delta \mu$ is found to be between 0.21 and 0.25 with very little deviation. For the last two events, the value of $\Delta \mu$ is found to be around 0.4 with greater uncertainties.


\section{Discussion}
\label{sec:discussion}

\subsection{Comparison of observational and simulation results}

Ray-tracing simulations are used to explore the variations of radio burst properties with respect to the variation of several parameters of the ambient plasma. As shown in \cite{kontar_etal_2019}, and reminded in this paper,  the time profile and directivity of the radio emission depends on the anisotropy factor introduced to account for anisotropic scattering of radio-waves on density fluctuations of the ambient plasma. Using the results of ray-tracing simulations with different anisotropy factor $\alpha$, we established a clear relation between the shape of the radio emission directivity, characterized by the parameter $\Delta \mu$, and the level of anisotropy. We also showed in this paper that simulation results suggest that this parameter $\Delta \mu$  depends only weakly on the radio emission frequency.

The directivity of type III radio bursts can only be determined from observations when simultaneous observations of type III radio bursts are available from several different vantage points. In this paper, we use observations at 4 different probes which are all located close to the ecliptic plane: we therefore limited our analysis in this plane and made the assumption that the radio sources were emitted in the ecliptic plane. 
For the first three events, the averaged values of $\Delta \mu$ lie between 0.21 and 0.25, which corresponds to anisotropy factors $\alpha$ between 0.36 and 0.44 in the simulations presented in section \ref{sec:simu}. For the last two events, the average value of $\Delta \mu$ is of 0.4, corresponding to an anisotropy factor of 0.6.

As shown in figure \ref{decaytime}, with an anisotropy factor as low as $\alpha = 0.30$, the decay times from the simulations are higher than the observed decay time by roughly a factor of 2. Given the uncertainties in the measurements and the approximations made in the simulations, this remains a good agreement. However, given that higher anisotropy factors will lead to increased decay times, it may be more difficult to explain the observed decay times for the last two events which anisotropy factor is believed to be around 0.6, and further investigation of the influence of the assumptions and other parameters in the simulation are necessary to completely explain these observations.
We also noted that anisotropic scattering of radio-waves introduced an "echo" which manifests in the light curves as a deviation from the exponential shape of the decay part of the curve. However, this deviation is subtle and visible after a time where the intensity drops roughly below 10\% of the peak intensity. This property might therefore be challenging to discern in the observations.

\subsection{Individual radio bursts properties compared to statistical results}

The present paper reports multi-spacecraft observations of single type III radio bursts, whose properties can be compared to the results of previous studies which examined such parameters statistically. 

As discussed in \ref{sec:obs}, the decay times of the radio bursts show similar of slightly lower values as decays times reported for STEREO and Parker Solar Probe observations \citep{krupar_etal_2020} or as compiled in \cite{kontar_etal_2019} from different studies. The decreasing trend of the decay time with increasing frequency is also observed on a single event basis. We showed here that this decay times does not vary significantly with the angle between the source and the point of observation (probe position). 

\begin{figure}
\includegraphics[width=0.98\linewidth]{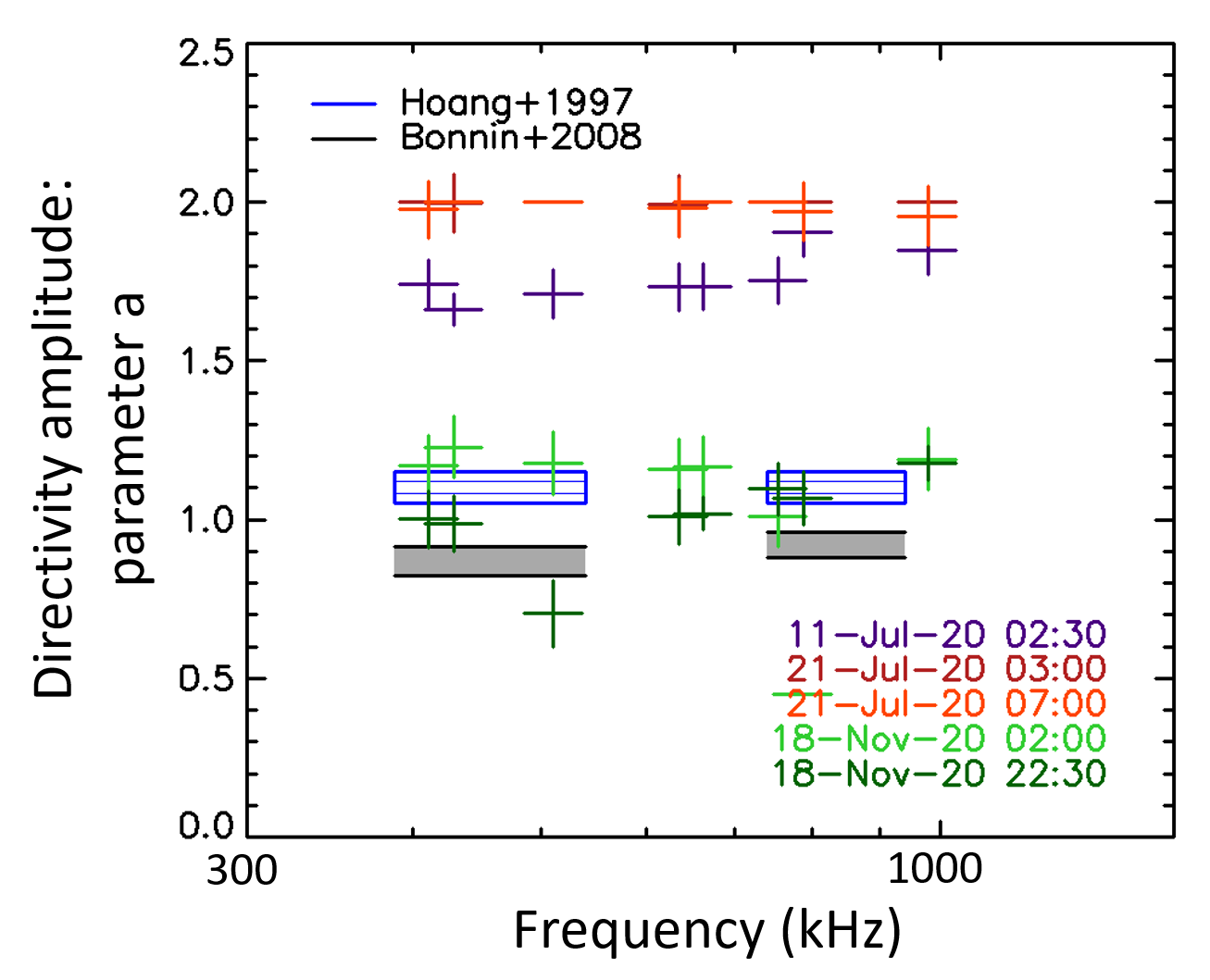}
\caption{Parameter $a$ describing the amplitude of the directivity, for the model described by equation \ref{eqn:bonnin}, and determined by a fit of this model to the peak fluxes for the 5 radio bursts selected, shown as coloured crosses. Previous results obtained in statistical studies of flux ratios of type III radio bursts by \cite{bonnin_etal_2008} and \cite{hoang_etal_1997} are also shown as squared boxes.}
\label{observation_a}
\end{figure}

On the other hand, the directivity profiles determined in this study for individual radio bursts can be compared to the directivity of radio bursts determined statistically on thousands of events as reported by \cite{bonnin_etal_2008,hoang_etal_1997}. In their study, the average directivity of radio burst was determined by fitting the distribution of flux ratios observed by two probes, for thousands of events. A slightly different model was then used for the directivity, in the form:
\begin{equation}
D(\varphi') = C_0 10^{(a ( \cos \varphi' -1))}
\label{eqn:bonnin}
\end{equation}
where $\varphi'$ is the angle between the source direction and the probe position, in a source-centered coordinate system.
In our analysis, we are considering the angles in a sun-centered system. However, since the frequency of our measurements is above 400 kHz, the radio sources remain very close to the Sun, and we consider that $\varphi' = \varphi$, with $\varphi$ being the difference between the probe longitude and the source longitude. In that specific case, the relation between parameters $a$ and $\Delta \mu$ is simple: $a = (\ln(10) \Delta \mu )^{-1}$.

We used this model described by equation \ref{eqn:bonnin} to fit the data, following the same method used to calculate the $\Delta \mu$ parameter, but this time to calculate the values of the parameter $a$ consistent with our observations, which is then compared to the values of this parameter found by the previous studies. The results are shown in figure \ref{observation_a}. The values of the parameter are of the same order of magnitude as the ones calculated in the statistical studies, which could be considered as averaged values. As it may be excepted, the values from a single event to another can vary significantly and differ from the average. We also note that the values of the parameter $a$ found by fitting the data verify the expected relation $a = (\ln(10) \Delta \mu )^{-1}$.

In the frequency range studied here, there is no obvious evolution of the directivity shape with frequency. \cite{bonnin_etal_2008} and \cite{hoang_etal_1997} reported a variation of the directivity with frequency which was significant at frequencies below 400 kHz. It can be noted that such a dependence is not observed in the ray-tracing simulations; however, at frequencies below 400 kHz, it is probable that we can no longer assume that the angles which are defined with respect to the centre of the Sun (as it is the case in the simulations) and those defined with respect to the source (as it is in these studies) can be considered as similar. This change in geometry could therefore account for the frequency dependence of the directivity profiles found in these past results.

We note that the directivity profile found for the last two events presented here are similar to the directivity found by the previous studies of \cite{bonnin_etal_2008} and \cite{hoang_etal_1997}, while the directivity profile for the first three event is narrower (with higher values of the parameter $a$). These observations demonstrate how the directivity profile can deviate from the averaged values determined by past studies. It is interesting to note that it is possible that the shape of directivity (i.e. the values of $\Delta \mu$ or $a$) seems to remain consistent in time: the two events most closely related in time (on July 21 2020) show almost identical directivity profiles. Since those events also share the same location, this can be interpreted as events for which the solar wind conditions remains the same. These results therefore demonstrate a potential for this type of analysis to probe the plasma and solar wind conditions at the sources of the radio emission in the inner heliosphere.


\section{Conclusion}

The launch of the Solar Orbiter and Parker Solar Probe missions, in combination with the STEREO and Wind missions, opened the opportunity for the simultaneous observations by four widely-separated spacecraft of solar events, such as type III radio bursts. Multi-spacecraft observations of these radio emissions are used here to study the directivity of single type III radio burst: in the past, directivity measurements could only be performed by a statistical analysis of radio flux ratios from many different bursts.
In the present study, we also provide an estimation of the source locations. As the data presented here was taken early in the Solar Orbiter mission, all measurements are performed close to the ecliptic plane and therefore, no information on the latitude of the radio sources can be reliably inferred, but as the Solar Orbiter mission continues, the satellite orbit will increase in inclination in regards to the ecliptic plane \citep{muller_etal_2020,zouganelis_etal_2020}, providing an opportunity to further study the directivity of the radio emissions in the 3-D space for individual events.

In this study, we looked at 5 different radio bursts and showed that while their directivity could be consistent with previous, averaged results, it could also vary significantly from these averaged values and from one event to another. 
The two closest event in time and space, on July 21 2020, show very similar directivity profiles, suggesting that the radio emission was emitted in similar solar wind conditions. This suggest, as one can expect, that the radio emission directivity will be significantly affected by the properties of the ambient plasma and solar wind.

Ray-tracing simulations of radio-wave propagation with anisotropic scattering on density fluctuations suggest that the directivity pattern of radio emission strongly depends on the ambient plasma conditions, and in particular on the anisotropy of the density fluctuations of the plasma. Multi-spacecraft radio emission diagnostics therefore enable us to characterise the anisotropy in the plasma density fluctuations. In the 400-1000 kHz range, we typically probe plasma in the range 5 to 20 solar radii from the Sun: in this range of distances, it will be possible to combine the type of observations presented in this paper with in-situ measurements by Parker Solar Probe at its closest approach, but also to get beyond the range explored in-situ by the probe. 


\begin{acknowledgements}
Solar Orbiter is a space mission of international collaboration between ESA and NASA, operated by ESA. 
Parker Solar Probe was designed, built, and is now operated by the Johns Hopkins Applied Physics Laboratory as part of NASA’s Living with a Star (LWS) program (contract NNN06AA01C). Support from the LWS management and technical team has played a critical role in the success of the Parker Solar Probe mission.
The Wind/WAVES investigation is a collaboration of the Observatoire de Paris, NASA/GSFC, and the University of Minnesota. 
S. Musset  is supported by an ESA Research Fellowship.
V. Krupar acknowledges the support by NASA under grants \texttt{18-2HSWO218\_2-0010} and \texttt{19-HSR-19\_2-0143}.
N. Chrysaphi thanks CNES for its financial support.
Data analysis was performed using IDL (Exelis Visual Information Solutions, Boulder, Colorado).
\end{acknowledgements}


\bibliographystyle{aa} 
\bibliography{zmabiblio} 
\end{document}